\documentclass[prb,aps,twocolumn,superscriptaddress]{revtex4-1}
\newcommand{\beq}{\begin{equation}}
\newcommand{\eeq}{\end{equation}}
\usepackage{graphics}
\usepackage{epsf}
\usepackage{amsmath}
\usepackage{amsfonts}
\usepackage{amssymb}
\usepackage{graphicx}
\usepackage{epsfig}

\begin{document}

\title{A high Neel temperature 5$d$ oxide:  NaOsO$_3$ }

\author{S. Middey}
\altaffiliation[Present address:]{ Department of Physics, University of Arkansas, Fayetteville, Arkansas 72701, USA }
\affiliation{Centre for Advanced Materials, Indian Association for the Cultivation of Science, Jadavpur, Kolkata-700032, India }
\author{Saikat Debnath}
\affiliation{S. N. Bose National Centre for Basic Sciences, JD-Block, Sector III, Salt Lake, Kolkata-700098, India}
\author{Priya Mahadevan}
\email[Corresponding author: ] {  priya.mahadevan@gmail.com}
\affiliation{S. N. Bose National Centre for Basic Sciences, JD-Block, Sector III, Salt Lake, Kolkata-700098, India}
\author{D. D. Sarma}
\affiliation{ Solid State and Structural Chemistry Unit, Indian Institute of Science, Bangalore-560012, India}
\affiliation{Council of Scientific and Industrial Research - Network of Institutes for Solar Energy (CSIR-NISE), New Delhi, India}
\begin{abstract}
The  origin of a high Neel temperature in a 5$d$ oxide, NaOsO$_3$, 
has been analyzed within the mean-field limit of a multiband Hubbard 
model and compared with the analogous 4$d$ oxide, SrTcO$_3$. 
Our analysis shows that there are a lot of similarities in both these
oxides on the dependence of the the effective exchange interaction strength
($J_0$) on the electron-electron interaction strength ($U$). However, the 
relevant value of $U$ in each system puts them in different portions
of the parameter space. Although the Neel temperature for NaOsO$_3$
is less than that for SrTcO$_3$, our results suggest that there could
be examples among other 5$d$ oxides which have a higher Neel temperature.
We have also examined the stability of the G-type antiferromagnetic 
state found in NaOsO$_3$ as a function of electron doping within GGA+U
calculations and find a robust G-type antiferromagnetic metallic state stabilized. The most surprising aspect of the doped results is the rigid band-like evolution of the electronic structure which indicates that the magnetism in NaOsO$_3$ is not driven by fermi surface nesting. 
\end{abstract}

\maketitle
PACS number(s): 75.10.-b, 75.47.Lx, 71.30.+h

\section{Introduction}

Metal-insulator transitions in strongly correlated systems, especially transition metal oxides still continue to be an important topic of research as new and unusual examples emerge~\cite{Imada_rmp}. The transitions in these systems have been understood within the Hubbard model~\cite{hubbard_nos} and arise from a competition between localizing effects controlled by the onsite Coulomb interaction strength $U$ and  electron delocalization
effects governed by the bandwidth, $W$. Within the Hubbard model, the system is insulating when $U/W$ is greater than 1, and a change of this ratio (across 1) by doping, external pressure etc. are routes by which one  can induce a metal-insulator transition. While a lot of examples exist among the 3$d$ oxides, there are very few insulating members among the 4$d$ and 5$d$ perovskite oxides. The larger spatial extent of the 4$d$ and the 5$d$ orbitals results in a larger bandwidth, and the additional screening effects present in these larger bandwidth oxides are expected to result in a reduced $U$. Consequently, the smaller $U/W$ ratio, expected to be less than 1, explains the metallic ground states observed in most cases. The few insulating examples that one finds there, require some other mechanism than the Mott transition to explain the ground state character. Recently, Shi {\it et al.} found a metal-insulator transition in NaOsO$_3$ at 410 K~\cite{NOO}, which has been observed as arising from the onset of antiferromagnetic order, thus the system is a perfect illustration of Slater insulator~\cite{Slater_nos}. Very recent neutron and X-ray scattering studies and optical spectroscopy further confirm the picture of a Slater transition in this compound~\cite{naoso3_condmat_nos,naoso3_infra}. As any long range magnetic order requires correlated electrons, it is rather a surprise, that the magnetic transition temperature of NaOsO$_3$ is so high (410 K). Local magnetic moments are usually very small in 5$d$ oxides as a result of the wide bands, so large ordering temperatures are quite unexpected.

The basic electronic structure of NaOsO$_3$ has been calculated before~\cite{NOO,naos_theory,pickett}, though the key question of why one has such  a high ordering
temperature in a 5$d$ oxide is still not well understood. Du {\it et al.}~\cite{naos_theory} recently showed that inspite of Os being a 5$d$ 
transition metal atom, spin-orbit interactions seem to have a weak effect on the band structure in NaOsO$_3$. The states near the 
fermi level seem to be strongly itinerant with a strong admixture of Os $d$ and O $p$ states suggesting that the appropriate
picture of magnetism seems to be an itinerant one. Further, they conclude by saying that the magnetic ordering as well as coulomb 
interactions play a role in making NaOsO$_3$ insulating. The nature of magnetism and the origin of the insulating state has been raised
by Jung {\it et al.}~\cite{pickett}. Considering a cubic ideal perovskite structure for NaOsO$_3$, they find that a small rotation of the OsO$_6$
octahedra is able to drive NaOsO$_3$ insulating as well as stabilize G-type antiferromagnetic ordering. Consequently, this should be
characterized by a spin spiral vector $q$=0 in contrast to a fermi surface nesting or hot spot driven magnetic transitions which
are characterised by a non-zero nesting vector. This makes the nature of the observed magnetic transition in NaOsO$_3$ even more 
puzzling, though Slater made  the observation in his seminal paper~\cite{Slater_nos} that the up and down spin electrons see different potentials,
and by that definition, even a transition characterized by a $q$=0 spin spiral vector should fit in as a Slater transition. 

In the present work, we have calculated the electronic and magnetic structure of NaOsO$_3$ within GGA based ab-initio electronic structure calculations. Both the G-AFM ground state  as well as its insulating nature are captured within these calculations. In order to understand the origin of the magnetic state, we perform further analysis within a multiband Hubbard model where the tight binding part is determined from a fitting of the ab-initio band structure. At a $U$ of just 0.8 eV, one is able to stabilize an insulating G-AFM state as the ground state. This scale of $U$ is far less than the bandwidth of $\sim$ 3 eV, suggesting that the scale of $U$ is irrelevant, and we have been able to capture the Slater insulator limit within our calculations. Varying $U$ and the charge transfer energy, one finds a wide range of parameter space for which the G type AFM solution is the ground state indicating the robustness of the magnetism. The origin of the magnetism can be traced back to the strong bonding-antibonding splitting arising from the interaction between nearest neighbor Os atoms. As the $t_{2g}$ levels are half-filled here, this 
energy gain is possible only for an antiferromagnetic configuration, while in the fully spin polarised limit the ferromagnetic configuration has no energy gain associated with it. Surprisingly, the small intraatomic exchange splitting and the large bandwidth aid to stabilize the magnetic state more as we pointed out in an earlier work~\cite{srtco3_own}, though the naive expectation would be that it destabilizes magnetism. This picture is further strengthened by our studies carried out as a function of doping.While the filling should be important in a model where magnetism is nesting-driven, in the present case we find that the G-type antiferromagnetism is robust even away from half-filling, though the ground state now becomes metallic. Examining
the exchange interaction strengths in the doped systems, we find an additive nature of the magnetic interactions emerges, with the partial filling of the Os $d$ $t_{2g}$ minority spin levels contributing a ferromagnetic component. This route is also suggestive of a clean route to obtain an antiferromagnetic metal as there is no strong electron-phonon coupling physics to 
be found here.

 The mechanism for magnetism proposed here for NaOsO$_3$ is similar what we had shown earlier~\cite{srtco3_own} for a wide variety of compounds which include SrTcO$_3$~\cite{srtco3_prl}, CaTcO$_3$~\cite{catco3_jacs} etc. though NaOsO$_3$ has much smaller $T_N$ (410 K) compared to SrTcO$_3$  (1023 K). This difference has been explained in the present work by a study of the variation of the effective exchange interaction strengths $J_0$ as a function of $U$. SrTcO$_3$ lies  at the peak of the curve, while NaOsO$_3$ lies in the rising part. This suggests that one may be able to find a different 5$d$ oxide which could have a $T_c$ higher than SrTcO$_3$.

\begin{figure}[b!]
\includegraphics[width=0.45\textwidth]{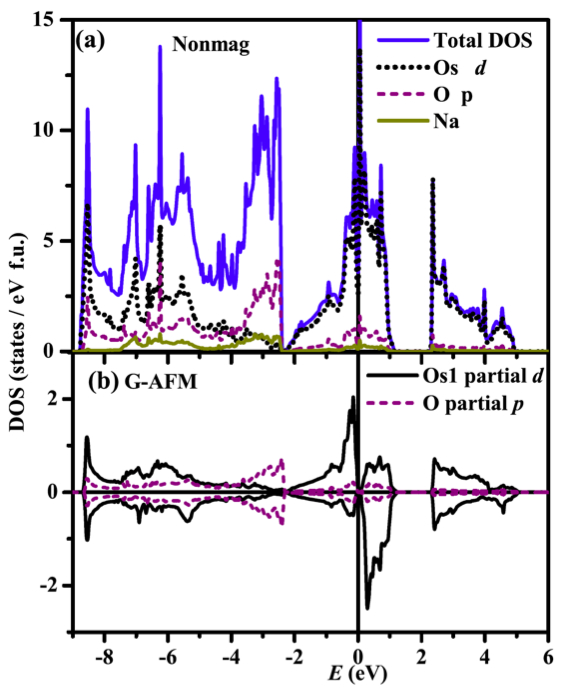}
\caption{(color online) Density of states of NaOsO$_3$ obtained from GGA calculation ($U$  =  0) (a) The total as well as partial density of states for nonmagnetic configuration. (b) The up (upper panel) and the down spin (lower panel) Os $d$ projected partial density of states for  G-AFM magnetic structure. The zero of energy corresponds to the Fermi energy.}
\end{figure}

\section{Methodology}
The electronic and magnetic structure of NaOsO$_3$ has been calculated within a plane wave pseudopotential 
implementation of density functional theory using PAW potentials~\cite{paw_potential} as implemented in VASP~\cite{vasp}. 
In addition to the GGA form for the exchange correlation functional, we have also considered electron-electron
interactions in the $d$ states of Os by considering a  $U$ on Os in the Dudarev implementation~\cite{dudarev_u_formalism} 
of the GGA + $U$ scheme. Different magnetic configurations~\cite{mag_def} were examined 
for different values of $U$. In our calculations we used a k-mesh of $6\times 6\times 6$ k-points and a cut off 
energy of 400 eV for the plane wave basis states. In these calculations, the lattice constants were kept fixed at 
the experimental values~\cite{NOO}, while the internal coordinates were optimized to minimize the total energy. We have used a sphere of radius 1.44 \AA \ about Os, 1.66 \AA \ about Na and 0.73 \AA \ about O for the calculation of the partial density of states as well the magnetic moment.
In order to understand the origin of the observed magnetic stability and its dependence on microscopic parameters $U$, 
charge transfer energy ($\Delta$), we carried out additional analysis in terms of a multiband Hubbard-like Hamiltonian 
given by equation 1. 

\begin{eqnarray}
H & = & \sum_{i,l,\sigma} \epsilon_p p_{il\sigma}^\dag p_{il\sigma} + 
\sum_{i,l,\sigma} \epsilon_d^l d_{il\sigma}^\dag d_{il\sigma} + 
\sum_{i,j,l_1,l_2,\sigma} (t^{l_1l_2}_{i,j,pp} p^\dag_{il_1\sigma} p_{jl_2\sigma} \nonumber\\
& & + h. c.) + \sum_{i,j,l_1,l_2,\sigma} (t^{l_1l_2}_{i,j,pd} d^\dag_{il_1\sigma} p_{jl_2\sigma} + h. c.)\nonumber\\
& & + \sum_{\alpha \beta \gamma \delta, \sigma_1 \sigma_2 \sigma_3 \sigma_4} U_{dd}^{\alpha \beta \gamma \delta} d^\dag_{\alpha \sigma_1}d^\dag_{\beta \sigma_2}d_{\gamma \sigma_3}d_{\delta \sigma_4}
\end{eqnarray}

where $d^\dag_{il\sigma}$($d_{il\sigma}$) creates (annihilates) an electron with spin $\sigma$ in the $l^{th}$ $d$-orbital 
on Os in the $i^{th}$ unit cell, $p^\dag_{im\sigma}$($p_{im\sigma}$) creates (annihilates) an electron  with 
spin $\sigma$ in the $m^{th}$ $p$-orbital on oxygen atom in the $i^{th}$ unit cell.
The details about solving this Hamiltonian can be found in Ref. ~\cite{srtco3_own}.  The total energies for different magnetic 
configurations were mapped onto a Heisenberg model (-$\frac{1}{2}\sum J_{ij}$ s$_{i}$.s$_{j}$) with 
first neighbor ($J_1$) as well as second neighbor ($J_2$) exchange interaction strengths. For the structures considered an effective 
exchange interaction strength $J_0$ given by 6$J_1$+ 12$J_2$ was determined in each case, as it could be related to the 
mean field $T_N$. 

The robustness of the results to electron doping has been examined by considering the 20 atom supercell of NaOsO$_3$ and replacing one or two Na atoms by Mg. This corresponds to an average occupancy of 3.25 or 3.5 at the Os site. Various magnetic configurations have been considered and we discuss the stability of the G-AFM state as a function of doping. Additionally, we have considered various non-collinear magnetic configurations in the vicinity of the G-AFM state
and determined their energies to see if any spin spiral magnetic solution had lower energy. Spin-orbit interactions are large in 5$d$ oxides and so we have calculated the stability of the G-AFM state including spin-orbit interactions to probe the role of spin-orbit interactions.

\begin{figure}[b!]
\includegraphics[width=0.45\textwidth]{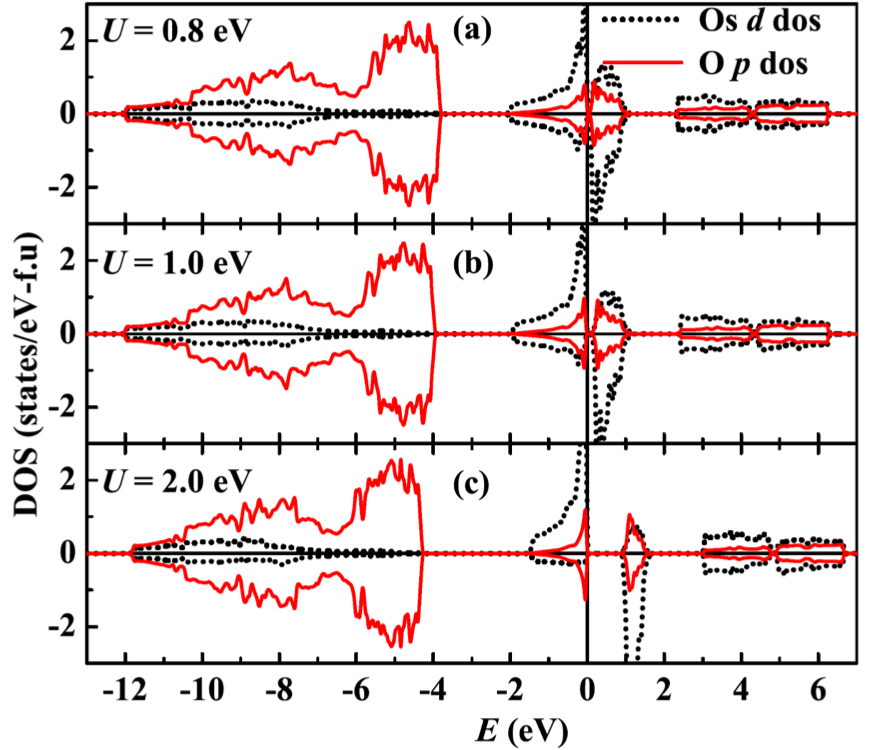}
\caption{(color online) The Up (upper panel) and down spin (lower panel) projected Os $d$ and O $p$ partial density of states for G-AFM spin configuration 
for $\Delta$=2~eV, $J_h$=0.1~eV as a function of $U$ from mean-field 
multiband Hubbard calculations for NaOsO$_3$. The zero of energy corresponds to the Fermi energy.}
\end{figure}

\section{Results and discussions}

The magnetic stabilization energies for different magnetic configurations 
calculated from our {\it ab-initio} calculations are given in Table I. In each case
the energies are referenced to the nonmagnetic configuration.
For the calculations in which we have $U$ = 0, we find that all the magnetic 
configurations except the G-AFM configuration converge to 
nonmagnetic solutions. The stability of the G-AFM state is found to be 
large with respect to the nonmagnetic configuration. 
The robustness of the G-AFM configuration vis-a-vis other magnetic 
configurations has been observed and discussed earlier in the context of 
SrTcO$_3$~\cite{srtco3_own}. This has been traced back to the large 
superexchange energy gain arising between nearest neighbor antiferromagnetic 
spins for a half-filling of the $t_{2g}$ orbitals on Os. As the G-type
antiferromagnetic configuration has all nearest neighbor spins
antiferromagnetically coupled, this energy gain from the superexchange
mechanism is the largest, and hence it is stabilized the most.
When $U$ is increased to 1 eV, one finds an increase in the stabilization
energy for the G-AFM state with respect to the closely competing C-type
antiferromagnetic configuration, which then begins to 
decrease as $U$ is increased further. The origin for this is clearer from
the model Hamiltonian calculations discussed later in the paper.

\begin{table}
\begin{center}
\caption
 {Stabilization energy (meV/f.u.) with respect to the nonmagnetic state from GGA+U calculations.}
\begin{tabular}{lcccc}
\hline
\hline
U(eV) & Ferro & A-AFM & C-AFM & G-AFM\\
\hline
0   & nonmag   & nonmag  & nonmag & -20 \\
1  & -7 & -11 & -50 & -127 \\
2 & -111 & -132 & -225 & -314\\
 3   & -295 & -415 & -521 & -587\\
\hline
\end{tabular}
\end{center}
\end{table}

The total and partial density of states (DOS) for the nonmagnetic solution 
is plotted in Fig.1(a).  The nonmagnetic solution is metallic, with 
the Fermi energy $E_F$ lying in the middle of  the $t_{2g}$ antibonding  states.
The large width (greater than 3 eV) of the $t_{2g}$ antibonding states 
and the significant admixture of Os $d$  and O $p$ states appear due to 
the larger spatial extent of the 5$d$ orbitals. Na acts as a perfect 
electron donor as suggested by the absence of significant Na partial 
density  states  (PDOS) near $E_F$. The up and down spin Os $d$ partial 
density of states for the G-AFM configuration is shown in Fig. 1(b). 
The G-AFM magnetic ordering modifies the DOS as shown in Fig. 1(b) by 
opening up a small gap at $E_F$. The magnetic moment on Os is found to be 
just 0.90 $\mu_B$,  close to the experimentally observed 
moment of 1 $\mu_B$~\cite{naoso3_condmat_nos}, but much smaller than 
the anticipated value of 3 $\mu_B$ which is expected for a $d^3$ 
configuration. The reduction in magnetic moment in this class of materials has sometimes been referred to as arising from spin fluctuation effects. The fact that we can capture this reduction in a mean field picture, that there is some other mechanism at play.
The reason for this is evident from the density of states 
which indicates a significant Os minority spin contribution 
in the density of states. 
Using a $U$ = 1 eV  on the Os $d$ states, we found that all the 
collinear magnetic configurations could be converged to magnetic
solutions. Considering the G-type AFM solution, one finds that 
the band gap for the is found to be 0.3 eV and the 
magnetic moment on Os is found to be 1.3 $\mu_B$.

\begin{figure}[t!] 
\includegraphics[width=0.45\textwidth]{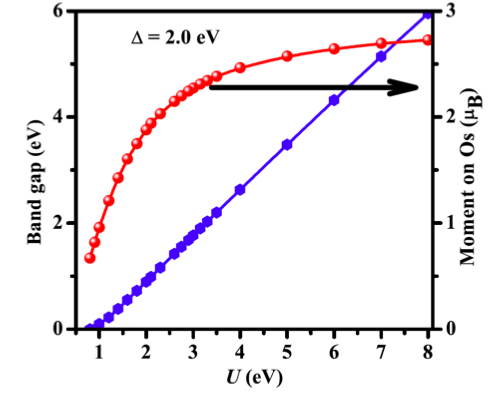}
\caption{(color online) The variation of the band gap (left axis) and the magnitude of magnetic moment of each Os atom (right axis) as a function of $U$ are plotted for $\Delta$ = 2.0 eV. }
\end{figure}

To understand the origin of magnetic ordering and the simultaneous 
metal-insulator transition, we set up a multiband Hubbard-like model 
for NaOsO$_3$ with a $U$ on the Os  $d$ states and calculated the 
electronic and magnetic ground states as a function of $U$ as
well as the charge transfer energy $\Delta$. The hopping matrix elements 
were parameterized in terms of the Slater Koster parameters 
$pd\sigma$, $pd\pi$, $sd\sigma$, $pp\sigma$ and 
$pp\pi$~\cite{slater_koster,mattheiss_prb}. A least squared error 
minimization procedure was used to estimate the best set of parameters 
entering the tight-binding part of the Hamiltonian that best fit the 
ab-initio band structure~\cite{priya_tb}. The bands with primarily 
Os $d$ character as well as the O $p$ nonbonding states were 
included in the fitting. The best-fit parameters were found to be
$sd\sigma$ = -3.65 eV, $pd\sigma$  =-3.75 eV, $pd\pi$ = 1.80 eV, 
$pp\sigma$ = 0.7 eV and $pp\pi$ = -0.15 eV and 
$\epsilon_d$  -  $\epsilon_p$ = 0.8 eV. The larger values of 
$pd\sigma$, $pd\pi$ compared to SrTcO$_3$, SrMnO$_3$~\cite{srtco3_own} 
are expected for a 5$d$ oxide and are due to the more 
extended nature of the 5$d$ orbitals.

The multiband Hubbard Hamiltonian for  NaOsO$_3$ has been solved 
for several values of $U$ using a mean-field decoupling scheme for the
four fermion term. 
We used a small value of 0.1 eV for the intraatomic Hund's exchange 
interaction strength and initially kept 
$\Delta$ (= $\epsilon_d$ - $\epsilon_p$ + 3$U$) fixed  at 2 eV.  
Again as observed in the case of the {\it ab-initio} calculations, here also 
we found that the G-type antiferromagnetic solution converged easily for even
a small value of $U$ of 0.8 eV, while other magnetic configurations 
considered did not converge to a magnetic ground state. Further, the G-AFM
configuration remained to be the ground state even as $U$ was increased.
In our calculations we have varied $U$ upto 8 ~eV. 
The Os $d$ and O $p$ partial density of states are plotted for the G-AFM 
configuration in Fig. 2  for several values of $U$. 
A bandgap is found to barely open up at a value of $U$ = 0.8 eV and 
this gradually increases 
as the value of $U$ is increased to 2~eV. 
Plotting the variation of the band gap ($E_g$) 
one finds an almost linear increase (Fig. 3). 
At lower values of $U$ one finds a significant contribution to the
Os $d$ occupied partial density of states from the minority spin channel.
This explains the strong reduction that one finds in the magnetic
moment from the value of 3 $\mu_B$ expected for a $d^3$ electronic
configuration. 
As the $U$ value is increased, the individual moment on each Os site 
increases and goes near saturation at larger $U$ values (Fig. 3).  
For the ferromagnetic case 
as well as the other antiferromagnetic configurations  where some neighboring spins are aligned ferromagnetically, the metal-insulator transition takes place at a larger value of $U$.  

\begin{figure} 
\includegraphics[width=0.45\textwidth]{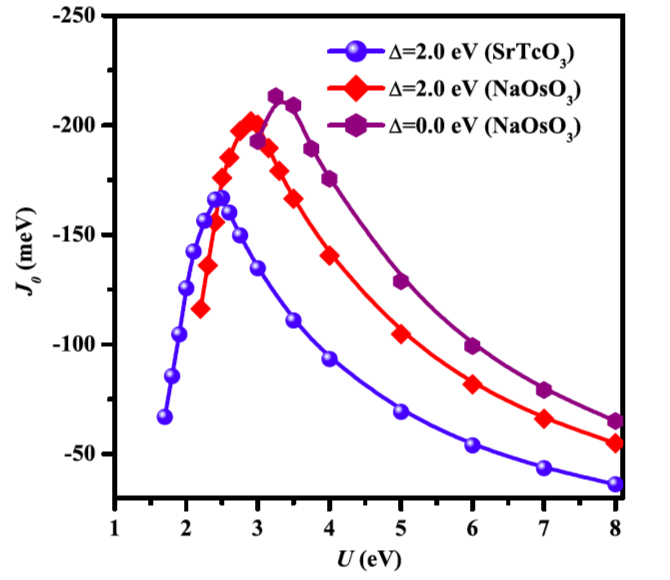}
\caption{(color online) Variation of effective exchange interaction 
strength ($J_0$) with $U$ for $J_h$ = 0.1 eV from mean field multiband 
calculations for $\Delta$=2 and 0 ~eV respectively. The hopping interaction
strengths for the tight-binding part of the Hamiltonian are taken for the
system shown in parentheses.}
\end{figure}

The next question we went on to address was whether the fact that SrTcO$_3$ 
had a higher $T_N$ than NaOsO$_3$ was indicative of a general trend
that 4$d$ transition metal oxides would have a higher ordering temperature 
than 5$d$ transition
metal oxides. 
To examine this we have calculated $J_0$ as a function of $U$ for the two systems. 
This is plotted in Fig. 4 for $J_h$=0.1 eV and $\Delta$=2 eV. 
For small values of  $U$, some of the magnetic configurations 
did not converge  to magnetic solutions. Consequently $J_0$ could not 
be estimated there. One aspect that is immediately apparent from Fig. 4
is that the variation of $J_0$ is similar in both systems.
The differences arise from the larger  
$p$-$d$ hopping strength that we have for NaOsO$_3$ as compared
with SrTcO$_3$. The peak of $J_0$  appears at $U$ = 2.5 eV and 
2.9 eV for SrTcO$_3$ and NaOsO$_3$ respectively. The value of $J_0$
at the peak is also much higher for NaOsO$_3$ than for SrTcO$_3$.
Now although we refer to the curve as representing variations for NaOsO$_3$
and SrTcO$_3$, these systems represent just one point on the curve,
characterised by the relevant $U$ for the system. 
The electron correlation strength in NaOsO$_3$ is expected 
to be much smaller than the $U$ value where $J_0$ has a peak. 
On the other hand, SrTcO$_3$ is located near the peak in $J_0$. 
This explains why NaOsO$_3$ has a smaller $T_N$. 
However, one could have another 5$d$ oxide which could have a much higher 
$T_N$ than SrTcO$_3$. To emphasize this point, we have varied $\Delta$ to 0 eV
and plotted 
the variation of $J_0$ in NaOsO$_3$.
A smaller $\Delta$ is expected to result in an increased 
superexchange interaction strength between nearest neighbor sites. 
This results in an increased $J_0$ compared to the
larger $\Delta$ solutions for the same value of $U$. The peak in $J_0$ is also
found to shift to a larger $U$. 
 
\begin{figure}
\includegraphics[width=0.45\textwidth]{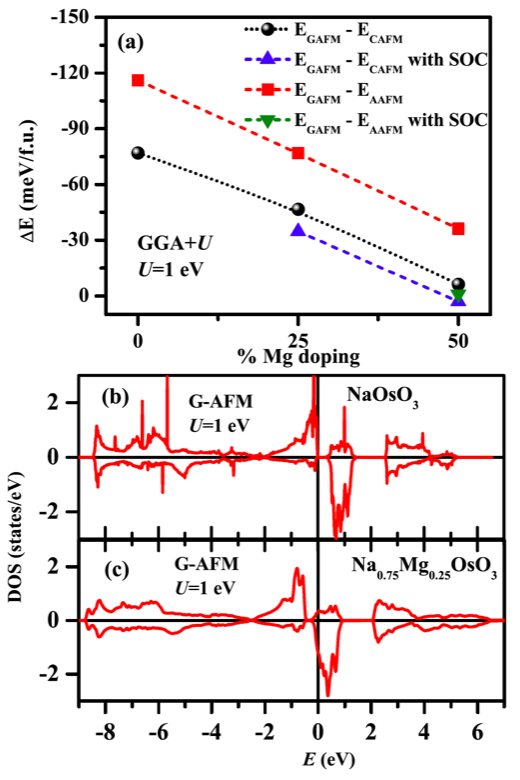}
\caption{(color online) (a) Variation of magnetic stabilization energy of 
G-AFM configuration with Mg doping ($U$ = 1 eV) from GGA+U calculations. 
Up (upper panel) and down spin (lower panel) Os $d$ projected 
partial density of states for (b) NaOsO$_3$ and (c) Na$_{0.75}$Mg$_{0.25}$OsO$_3$
at $U$= 1~eV in the G-AFM configuration. }
\end{figure}

Fermi surface nesting of a bipartite lattice~\cite{nesting} is generally 
used to explain the insulating ground state that emerges for the G-AFM 
magnetic structure. A small doping should take one away from the point where a robust G-AFM magnetic solution is found and so one does not expect any 
magnetic solution to be stable. In order to examine this aspect, monovalent  
Na of NaOsO$_3$ has been replaced partially by divalent Mg and the energy 
of various magnetic configurations have been calculated within $ab-initio$   
GGA+$U$ calculations. The lattice constants were fixed at the experimental 
value of NaOsO$_3$, 
while the internal coordinates were 
optimized. The stabilization energy of the lowest energy configuration   
(G-AFM ) has been plotted in Fig. 5(a) with respect 
to other antiferromagnetic configurations  (C-AFM, A-AFM)
as a function of Mg doping. 
Interestingly, we find an almost linear variation of the stabilization 
energies with doping. 
With an increase in Mg doping i.e. electron doping, 
the energy difference decreases and for 50\% doping case, 
the C-type and G-type antiferromagnetic configurations are 
almost degenerate. 
Spin-orbit interactions are believed to be sizable in 5$d$ oxides and 
could be of the same strength as the intra-atomic 
exchange interaction strength~\cite{ir1,ir2}. Consequently, we went 
on to examine if including spin-orbit interactions modified 
the conclusions we had in the absence of including spin-orbit 
interactions. A similar trend in the
magnetic stabilization energies was found even in the presence of spin-orbit 
interactions with the G-AFM spin configuration still found to be the
ground state for the 25\% doped case.

\begin{figure}
\includegraphics[width=0.45\textwidth]{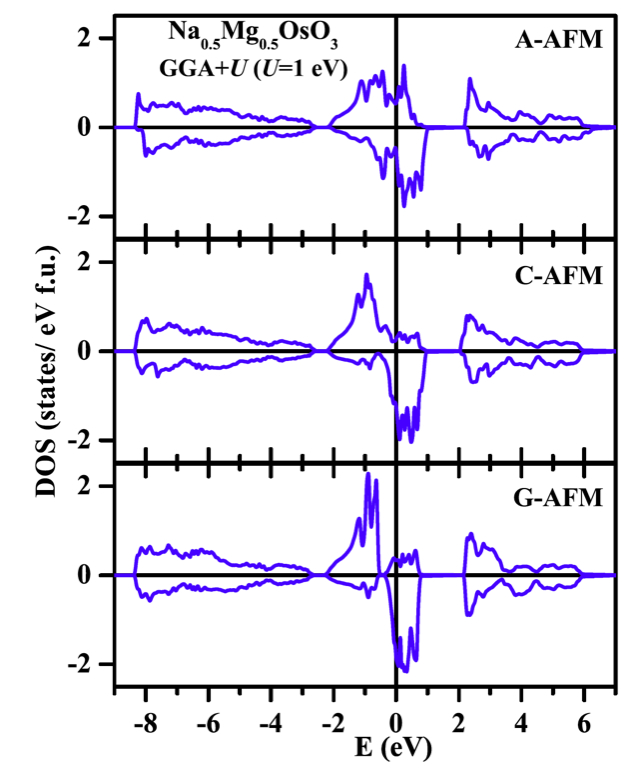}
\caption{(color online) Up (upper panel) and down spin (lower panel) Os $d$ projected 
partial density of states for Na$_{0.50}$Mg$_{0.50}$OsO$_3$ in
various magnetic configurations from GGA+U calculations. The zero
of energy represents the fermi energy.}
\end{figure}

The up and down spin Os $d$ partial density of states 
within the {\it ab-initio} calculations for $U$=1 are plotted in 
Figs. 5(b) and (c) for NaOsO$_3$ as well as the 25$\%$ doped case 
in the G-AFM magnetic structure. Surprisingly, one finds an
almost rigid band evolution of the density of states in the doped system.
A similar magnitude of the energy gap between the majority spin and 
minority spin $t_{2g}$ states is found with all other aspects of the 
electronic structure remain the similar. The Mg doping has now resulted in 
a shift of the fermi level into the minority spin $t_{2g}$ states and
we have a rare occurrence of an antiferromagnetic metallic state.

It is also interesting to note that while the $t_{2g}$ states show a large exchange splitting, the $e_g$ states which are at higher energies show hardly any exchange splitting. This is because the intra-atomic exchange splitting in these systems is small, and whatever exchange splitting one finds for the $t_{2g}$ states, emerges from the superexchange interactions, primarily from nearest neighbor Os atoms. This effect is even more evident in Fig. 6 where we have plotted the up and down spin Os $d$ partial density of states for different magnetic configurations for 50\% Mg doped NaOsO$_3$. The number of antiferromagnetic neighbors increases from A-AFM configuration to the C-AFM configuration to the G-AFM configuration. The exchange splitting is also found to increase with the number of antiferromagnetic neighbors contributing to the superexchange pathways. While there is no band gap between the majority spin and the minority spin $t_{2g}$ states for both A and C-AFM configurations, a small band gap exists for the G-AFM
magnetic structure, again emerging for the same reasons.

\begin{figure}
\includegraphics[width=0.45\textwidth]{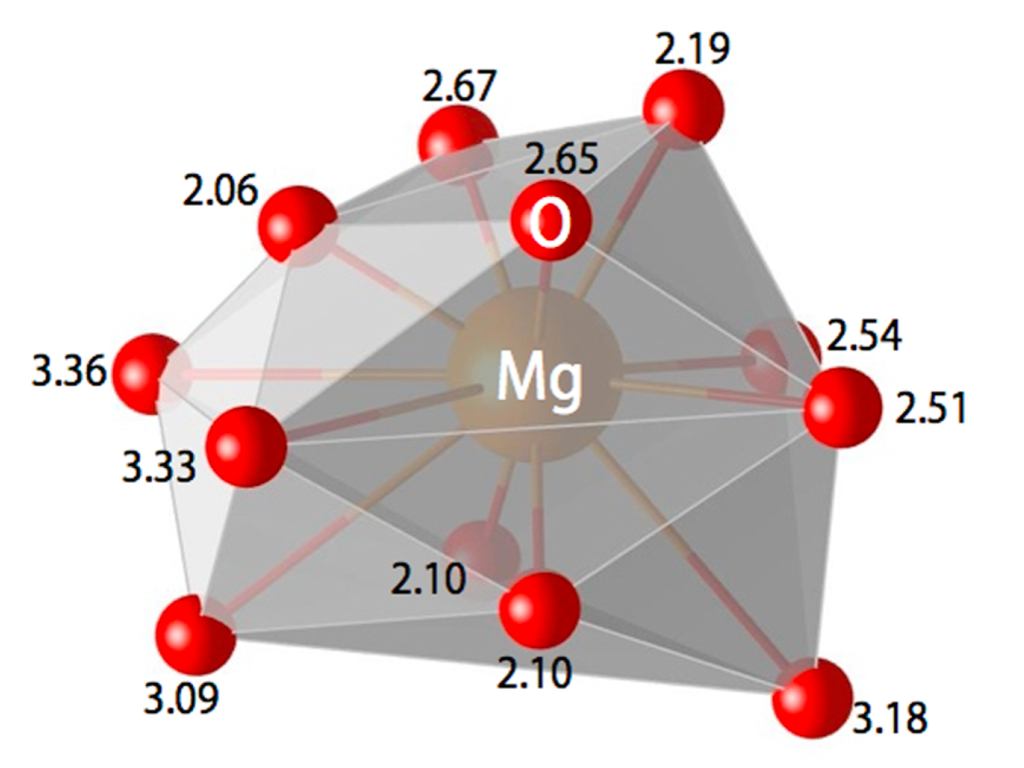}
\caption{(color online) Distortion in Mg-O bond lengths for 3.125\% Mg doped case.}
\end{figure}

In order to examine the stability of this antiferromagnetic state to the formation of polarons, we have constructed a 160 atom unit cell of NaOsO$_3$ and in that cell we have replaced one Na atom with Mg atom. We find that there are strong lattice relaxations of the oxygen atoms which form the first shell of neighbours around Mg. We show that the first shell of Mg-O atoms. Significant distortions are found with the shortest bond length equal to 2.06 \AA \ as shown in Fig. 7. However there is a very small effect on the Os-O bond lengths. Hence no strong electron-phonon physics are operative here. Hence doping electrons into a 5d oxide does not have the problems one encounter with 3d oxides such as the manganites \cite{polaron}  where polaron formation has been reported. These systems are suggestive of a facile route to a G-type antiferromagnetic metallic solution.

In Fig. 5(a) we have reported the magnetic stabilization energies for few collinear configurations. The immediate question which arises is whether we have probed enough magnetic configurations as the doped systems could have spin spiral configurations being favoured. We have therefore considered excitaions about the G-type antiferromagnetic configuration and examined various spin spiral configurations characterized by the spin spiral vector q as plotted in Fig. 8. The G point which has the lowest energy corresponds to the G-type antiferromagnetic solution. This is metallic for the 25\% Mg doped NaOsO$_3$ while it is insulating for the undoped case. Both these solutions establish that electron doping indeed stabilizes G-type antiferromagnetic metallic state in this system.

\begin{figure}
\includegraphics[width=0.5\textwidth]{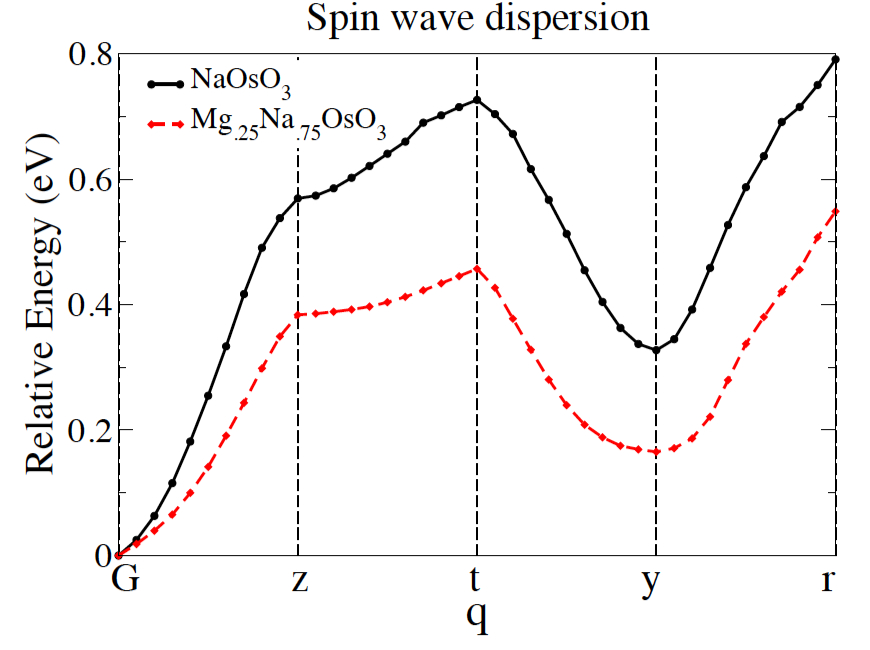}
\caption{ (color online) Spin wave dispersion along different symmetry directions for NaOsO$_3$ and 25\% Mg doped case. The zero of energy corresponds to energy of G-AFM state for both cases.}
\end{figure}


\section{Conclusions}
We have studied the  origin of a high Neel temperature in 5$d$ 
oxide NaOsO$_3$  within the mean-field limit of a multiband Hubbard 
model and compared that with 4$d$ oxides SrTcO$_3$.  Both these compounds 
show very similar trends in the variation of  the exchange interaction 
strengths.  The relevant strength of $J_0$ for NaOsO$_3$ places it in 
a region where $J_0$ increases with an increase in $U$ whereas SrTcO$_3$ 
is located near the peak in $J_0$. Examining the relevant 
parameter space for 5$d$ oxides, our analysis also suggests
that a higher $T_N$ is possible in a 5$d$ oxide by a suitable tuning of the
parameters. Additionally, we have examined the stability of the G-type antiferromagnetic state on electron doping. Not only do we obtain a rigid band-like evolution of the electronic structure on doping, but also we find that the G-type antiferromagnetism remains robust even when the system becomes
metallic. These systems throw up a rare occurrence of an antiferromagnetic metal. Important insights into the nature of magnetism of the parent compound can be inferred from the doping dependent studies.

\section{Acknowledgement}
SM and SD thank CSIR, India for fellowship. DDS and PM thank the Department of Science and
Technology, India.


\begin{thebibliography}{99}
\bibitem{Imada_rmp}M. Imada, A. Fujimori , and Y. Tokura, Rev. Mod. Phys {\bf 70}, 1039 (1998).
\bibitem{hubbard_nos} J. Hubbard, Proc. R. Soc. {\bf A276}, 238 (1963).

\bibitem{NOO}Y. G. Shi, Y. F. Guo, S. Yu, M. Arai, A. A. Belik, A. Sato, K. Yamaura, E. Takayama-Muromachi, H. F. Tian, H. X. Yang, J. Q. Li, T. Varga, J. F. Mitchell, and S. Okamoto, Phys. Rev. B {\bf 80}, 161104 (2009).
\bibitem{Slater_nos} J. C. Slater, Phys. Rev. {\bf 82}, 538 (1951).
\bibitem{naoso3_condmat_nos}S. Calder,  V. O. Garlea,  D. F. McMorrow,  M. D. Lumsden,  M. B. Stone,  J. C. Lang,  J.-W. Kim, J. A. Schlueter, Y. G. Shi, K. Yamaura, Y. S. Sun, Y. Tsujimoto, and A. D. Christianson, Phys. Rev. Lett. {\bf 108}, 257209 (2012).
\bibitem{naoso3_infra}I. Lo Vecchio,	A. Perucchi, P. Di Pietro, O. Limaj, U. Schade, Y. Sun, M. Arai, K. Yamaura, and S. Lupi,  Sci. Rep. {\bf 3}, 2990 (2013).
\bibitem{naos_theory}Y. Du, X. Wan, L. Sheng, J. Dong, and S. Y. Savrasov, Phys. Rev. B {\bf 85}, 174424 (2012).
\bibitem{pickett}M.-C. Jung,, Y.-J. Song, K.-W. Lee, and W. E. Pickett, Phys. Rev. B  {\bf 87}, 115119 (2013).
\bibitem{srtco3_own}   S. Middey, A. K. Nandy, S. K. Pandey, P. Mahadevan and D.D. Sarma, Phys. Rev. B  {\bf 86}, 104406 (2012).
\bibitem{srtco3_prl} E. E. Rodriguez, F. Poineau, A. Llobet, B. J. Kennedy, M. Avdeev, G. J. Thorogood, M. L. Carter, R. Seshadri, D. J. Singh, and A. K. Cheetham, Phys. Rev. Lett. {\bf 106}, 067201 (2011).
\bibitem{catco3_jacs}M. Avdeev, G. J. Thorogood, M. L. Carter, B. J. Kennedy, J. Ting, D. J. Singh, and K. S. Wallwork, J. Am. Chem. Soc. {\bf 133}, 1654 (2011).



\bibitem{paw_potential}P. E. Bl$\ddot{o}$chl, Phys. Rev. B \textbf{\textbf{50}}, 17953 (1994); G. Kresse and D. Joubert, \emph{ibid.} \textbf{59}, 1758 (1999).
\bibitem{vasp}G. Kresse, and J. Furthm\"{u}ller, Phys. Rev. B. \textbf{54} 11169 (1996); Comput. Mater. Sci. \textbf{6}, 15 (1996).
\bibitem{dudarev_u_formalism} S. L. Dudarev, G. A. Botton, S. Y. Savrasov, C. J. Humphreys, and A. P. Sutton, Phys. Rev. B \textbf{57}, 1505 (1998).

\bibitem{mag_def} The spin arrangements for different antiferromagnetic  configurations can be found in Ref. ~\cite{srtco3_own}.
\bibitem{slater_koster} J. C. Slater, and G. F. Koster, Phys. Rev. \textbf{94}, 1498 (1954).
\bibitem{mattheiss_prb}L. F. Mattheiss, Phys. Rev. B \textbf{2}, 3918 (1970).

\bibitem{priya_tb}P. Mahadevan, N. Shanthi, and  D. D. Sarma, Phys. Rev. B  {\bf 54}, 11199 (1996).
\bibitem{nesting}J. E. Hirsch, Phys. Rev. B {\bf 31}, 4403 (1985).
\bibitem{ir1} D. Haskel, G. Fabbris, Mikhail Zhernenkov, P. P. Kong, C. Q. Jin, G. Cao, and M. van Veenendaal, Phys. Rev. Lett. {\bf 109}, 027204 (2012)
\bibitem{ir2}B. J. Kim, Hosub Jin, S. J. Moon, J.-Y. Kim, B.-G. Park, C. S. Leem, Jaejun Yu, T. W. Noh, C. Kim, S.-J. Oh, J.-H. Park, V. Durairaj, G. Cao, and E. Rotenberg, Phys. Rev. Lett. {\bf 101}, 076402 (2008).
\bibitem{polaron} M. B. Salamon, and M. Jaime, Rev. Mod. Phys. {\bf 73}, 583 (2001).

\end{thebibliography}
\end{document}